\documentclass[8pt,pdflatex,sn-mathphys-num]{sn-jnl}

\usepackage{graphicx}%
\usepackage{multirow}%
\usepackage{amsmath,amssymb,amsfonts}%
\usepackage{amsthm}%
\usepackage{mathrsfs}%
\usepackage[title]{appendix}%
\usepackage{xcolor}%
\usepackage{textcomp}%
\usepackage{manyfoot}%
\usepackage{booktabs}%
\usepackage{algorithm}%
\usepackage{algorithmicx}%
\usepackage{algpseudocode}%
\usepackage{listings}%
\usepackage{caption}

\theoremstyle{thmstyleone}%

\theoremstyle{thmstyletwo}%

\theoremstyle{thmstylethree}%

\begin{document}

\title[Article Title]{Highly Efficient and Broadband Optical Delay Line towards a Quantum Memory}

\author[1]{\fnm{Yu} \sur{Guo}}\email{guoyu04@u.nus.edu}
\equalcont{These authors contributed equally to this work.}

\author*[1]{\fnm{Anindya} \sur{Banerji}}\email{cqtab@nus.edu.sg}
\equalcont{These authors contributed equally to this work.}

\author[1]{\fnm{Jia Boon} \sur{Chin}}\email{jbchin@u.nus.edu}

\author[1,2,3]{\fnm{Arya} \sur{Chowdhury}}\email{e0540843@u.nus.edu}

\author[1,3]{\fnm{Alexander} \sur{Ling}}\email{alexander.ling@nus.edu.sg}

\affil[1]{\orgname{Centre for Quantum Technologies, National University of Singapore}, \orgaddress{\street{S15, Science Drive 2}, \postcode{117543}, \country{Singapore}}}

\affil[2]{\orgname{Thales Research and Technology, 12 Ayer Rajah Crescent, Singapore 139941}}

\affil[3]{\orgname{Department of Physics,  National University of Singapore, 2 Science Drive 3,  Singapore, 117551}}

\abstract{We demonstrate a high-efficiency, free space optical delay line utilizing a nested multipass cell architecture. This design supports extended optical paths with low loss, aided by custom broadband dielectric coating that provides high reflectivity across a wide spectral bandwidth. The cell is characterized using polarization-entangled photon pairs, with signal photons routed through the delay line and idler photons used as timing reference. Quantum state tomography performed on the entangled pair reveals entanglement preservation with a fidelity of $99.6(9)\%$ following a single-transit delay of up to $687$~ns, accompanied by a photon retrieval efficiency of $95.390(5)\%$. The delay is controllable and can be set between $1.8$~ns to $687$~ns in $\sim12.6$~ns increments. The longest delay and wide spectral bandwidth result in a time-bandwidth product of $3.87\times 10^7$. These results position this delay line as a strong candidate for all-optical quantum memories and synchronization modules for scalable quantum networks.}

\keywords{optical delay line, quantum memory, quantum network}



\maketitle

\section{Introduction}\label{sec:Introduction}
Optical quantum communication networks promise secure communication, enhanced sensing capabilities, and distributed quantum computing \cite{kimble2008quantum, wehner2018quantum}. A critical challenge in realizing functional quantum networks is the development of reliable memory systems that can synchronize distant nodes \cite{simon2010quantum, bussieres2013prospective}. These memories must efficiently store and retrieve quantum states while preserving their coherence and entanglement properties \cite{heshami2016quantum}.

Well-established approaches to quantum memory have predominantly focused on light-matter interaction. Some of the commonly studied platforms include atomic ensembles \cite{fleischhauer2002quantum, reim2011single}, rare-earth-doped crystals \cite{zhong2015optically, afzelius2009demonstration, zhong2017science}, and spin defects \cite{shim2013pra, Jelezko2004prl, Dutt2007science, Neumann2008science, abobeih2019atomic, bradley2019ten}. Storage times ranging from milliseconds to hours have been achieved. Despite such advances, continued challenges are encountered related to low efficiencies, limited wavelength and bandwidth tunability, or the need for strict temperature control \cite{elizabeth2023, hedges2010efficient, zheng2022room}.

All-optical quantum memories \cite{lvovsky2009optical} present an alternate approach where photonic quantum information is stored directly in optical paths \cite{saglamyurek2011broadband, pittman2002probabilistic, Pittman2002, PremKumar2024, saglamyurek2015fiber, bustard2022prl, xiong2016nature, wolters2017prl, bouillard2019prl, makino2016scienceadvances, meyerscott2022prl}. One such example is the optical delay line based on fiber loops \cite{pittman2002probabilistic, Pittman2002, PremKumar2024, saglamyurek2015fiber, bustard2022prl, xiong2016nature}, which has demonstrated quantum state preservation but suffers from some of the same problems, e.g. bandwidth limitations, as the other platforms \cite{pittman2002probabilistic, Pittman2002, PremKumar2024}. 

Multipass cells \cite{herriott1964appliedoptics, herriott1964off, robert2002vacuum} are another type of optical delay line. A typical design is the Herriott cell, which consists of two spherical mirrors arranged coaxially to create multiple reflections of a beam. All multipass cells have low dispersion, can operate at room temperature, and have relaxed wavelength and bandwidth requirements. These properties make them promising candidates as all-optical quantum memories \cite{kaneda2015optica, kaneda2019scienceadvances, Arnold2024}. 

Conventional Herriott cells \cite{herriott1964off} restrict the beam reflection spots to elliptical or circular patterns. This underutilizes the mirror surface area, limiting the optical path length. To increase the optical path length, modifications to the cell design, such as introducing a perturbing mirror \cite{herriott1964appliedoptics} or segmented mirrors \cite{robert2002vacuum}, have been explored to generate more complex patterns utilizing more of the mirror surface. These solutions can be difficult to control and model accurately. 

An optical delay line based on nested multipass cells that was proposed in \cite{xue2023generalized} addresses the above-mentioned limitations. The reflection spots trace out a concentric circular pattern while adhering to the re-entrant condition of the Herriott cell. This results in a stable configuration while ensuring optimal utilization of the mirror surface, leading to longer optical paths. However, the application of the design in quantum memories remains unexplored.

We present a high-efficiency nested Herriott cell that can act as a building block for an all-optical quantum memory. The cell operates at room temperature with broad bandwidth compatibility and controllable delay times. It reaches a transmission efficiency of $95.390(5)~\%$ for a delay time of $687$~ns in a single transit. It can preserve polarization-entangled states with a fidelity of $99.6(9)~\%$ for the same delay. The high retrieval efficiency, together with a high fidelity and time-bandwidth product of approximately $3.87 \times 10^7$, make it particularly suitable for synchronization applications in quantum networks when coupled with fast optical switches \cite{pengfei2024}.

\section{Results}\label{sec:Result}

\subsection{Controllable delay and broadband high efficiency performance}

\begin{figure}[h]
\begin{center}
\includegraphics[width=\textwidth]{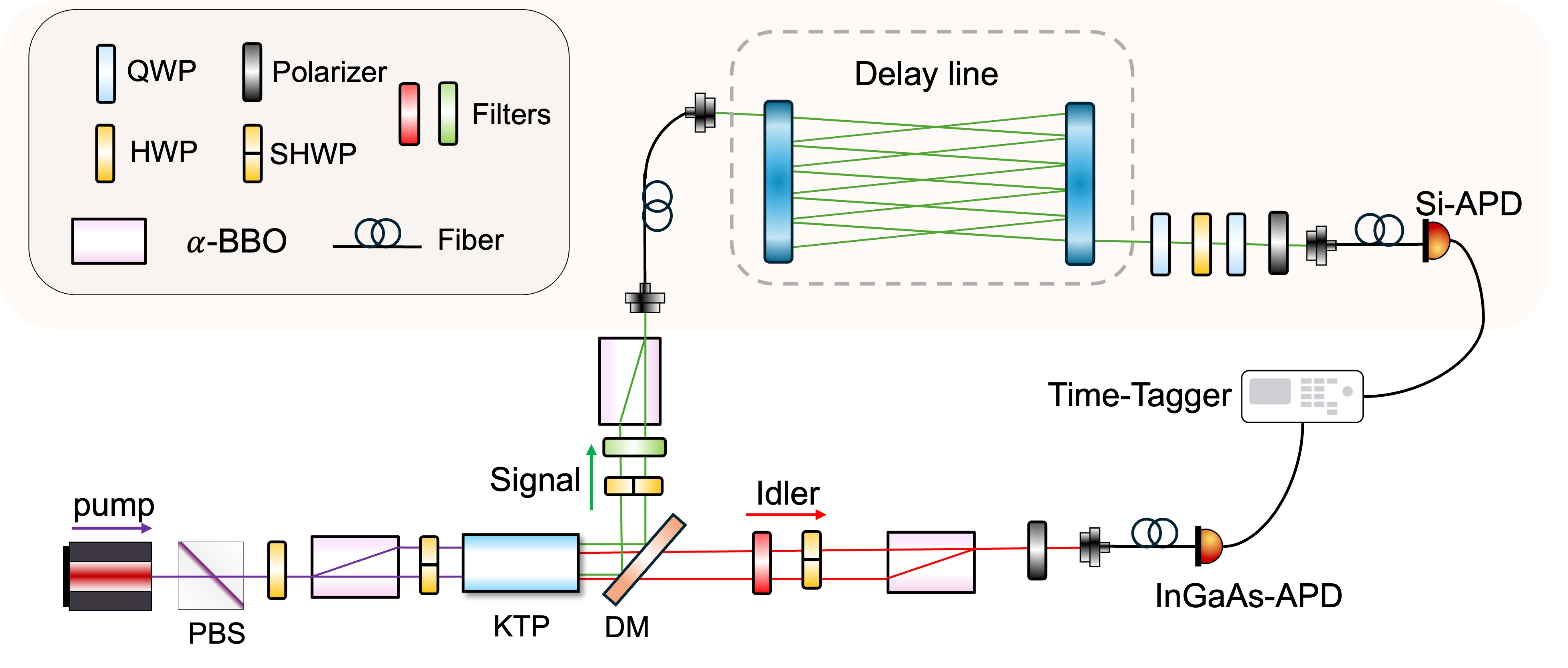}
\caption{Experimental setup. An entangled photon-pair source produced non-degenerate entangled photons at $562.7$~nm (signal) and $1454.9$~nm (idler). The idler photons were detected at source using an InGaAs avalanche photodiode (APD) while the signal photons were detected after passing through the delay line using a Si APD. Waveplates were used for compensating the polarization rotation induced by the fiber. PBS: polarization beam splitter; KTP: $\mathrm{KTiOPO_{4}}$ crystal; DM: dichroic mirror; $ \mathrm{\alpha-BBO}$: alpha-barium borate crystal; HWP: half waveplate; QWP: quarter waveplate; SHWP: segmented half waveplate. }
\label{fig:setup}
\end{center}
\end{figure}

The experimental setup is illustrated in Fig. \ref{fig:setup} and consists of three key components: an entangled photon-pair source, the optical delay line, and the detection system. The source generates polarization-entangled Bell states of the form
\begin{equation}
    \label{Eq:Bell_State}
     \vert \phi^+\rangle = \frac{1}{\sqrt{2}}\left(\vert H_s\rangle \vert H_i \rangle + \vert V_s\rangle \vert V_i \rangle\right),
\end{equation}
where the subscripts $s$ and $i$ denote signal and idler photons, respectively. The photon pairs are produced in a type-II spontaneous parametric down-conversion (SPDC) process via non-critical phase-matching \cite{JiaBoon2025}, pumped by a $405.7$~nm continuous-wave laser. The resulting non-degenerate photon pairs are centered at $562.7$~nm (signal) and $1454.9$~nm (idler).

The signal from the SPDC source is routed into the nested multipass cell, where it undergoes multiple reflections. These reflections collectively define the optical delay. The reflection spots trace out stable, concentric ring patterns on the mirrors. 

\begin{figure}[h!]
\centering
\includegraphics[height=11cm,width=0.8\textwidth]{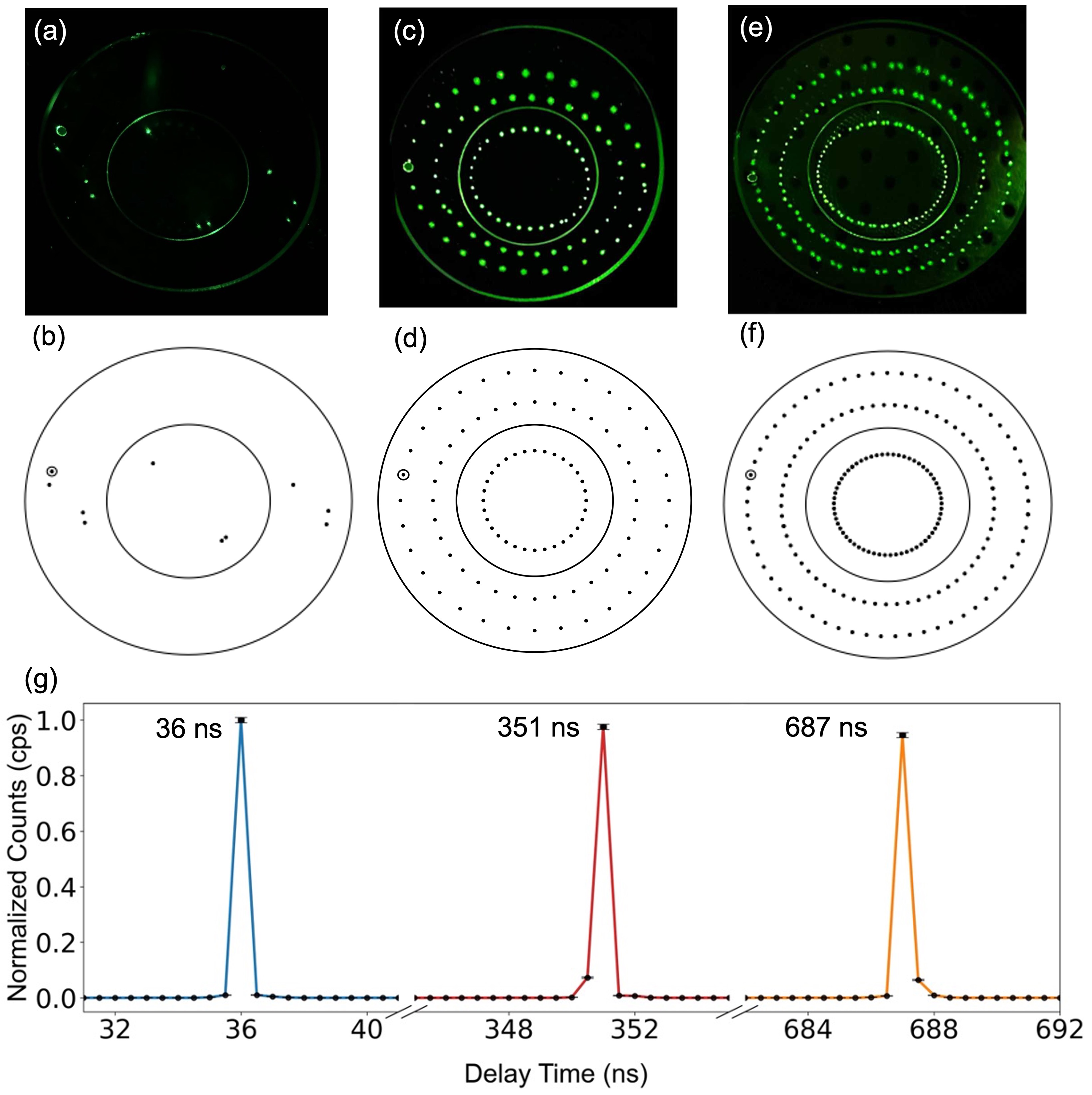}
\caption{Different configurtaion of reflection spot pattern and delay. (a), (c) and (e) are photographs of the exit mirror showing the concentric ring pattern with 9, 96 and 189 reflections. The total number of reflections in the cell is double these numbers. For visual clarity, these photographs were obtained using a $532.0$~nm laser. Their simulated spot patterns are shown in (b), (d) and (f).  Each pattern, and its delay time, was obtained by rotating the output mirror to reposition the exit pupil along the outermost ring.  (g) is corresponding delay histograms for above configurations.}
\label{fig:spotspattern}
\end{figure}

Rotation of the exit mirror to reposition the pupil along the outermost ring allows discrete control of the number of reflections, and hence the delay time, while preserving the spot geometry. Representative reflection patterns corresponding to different delays are shown in Fig. \ref{fig:spotspattern} alongside their simulated counterparts.  

The difference in time of arrival between the idler photon (detected directly) and the delayed signal photon was measured with a temporal resolution of $450$~ps, limited by the timing jitter of the single photon detectors. Systemic delays from electronics and additional optical path were calibrated and subtracted. Delay histograms for the three spot patterns  are shown in Fig. \ref{fig:spotspattern}(g). Their corresponding delay times are $36$~ns, $351$~ns, and $687$~ns.

The system maintains high retrieval efficiency across all delay times. As shown in Table.~\ref{tab:efficiency}, the delay line achieves an efficiency of $99.680(5)\%$ for $36$~ns delay, with an estimated per-reflection reflectance of $99.982(7)\%$. At the maximum delay of $687$~ns, the retrieval efficiency reaches $95.390(5)\%$, corresponding to a per-reflection reflectance of $99.988(7)\%$. 

\begin{table}[h]
\centering
\begin{tabular}{|c|c|c|c|}
\hline
No. of Spots& Delay time(ns)  & Efficiency(\%) & Reflectance (\%)\\
\hline
18 & 36 & 99.680(5) & 99.982(7)\\
\hline
42 & 77 & 99.370(2) & 99.985(8)\\
\hline
192 & 351 & 98.060(3) & 99.989(8)\\
\hline
204 & 370 & 97.480(2) & 99.988(7)\\
\hline
228 & 411 & 96.770(1) & 99.986(1)\\
\hline
378 & 687 & 95.390(5) & 99.988(9)\\
\hline
\end{tabular}
\caption{Table of the measured efficiencies and estimated reflectances per-reflection for 6 different settings.}
\label{tab:efficiency}
\end{table}

\begin{figure}[h!]
\centering
\includegraphics[width=0.75\textwidth]{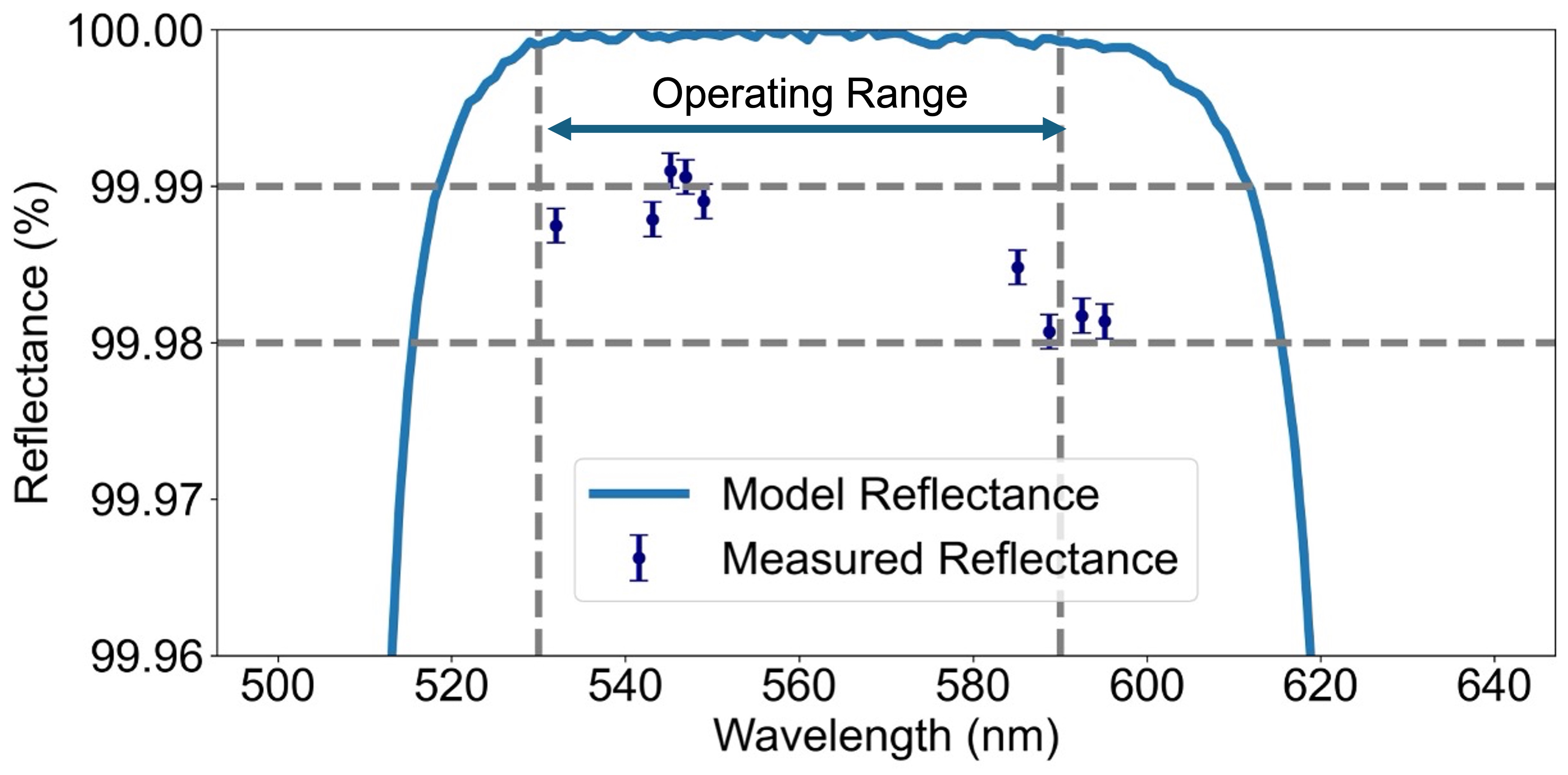}
\caption{Comparison of the estimated reflectance of the mirrors with the manufacturer supplied data. The simulated reflectance curve (solid line) was provided by the mirror manufacturer.  The estimated reflectances were obtained by measuring the retrieval efficiency over the wavelength range of $532–595$~nm, with a delay of $370$~ns corresponding to 204 spots. The consistent lower reflectances were a result of scattering through floating dust particles and slight attenuation in free space. The area enclosed by the vertical dashed lines corresponds to the operational wavelength range.  }\label{fig:Measured_Reflectance}
\end{figure}

In addition to controllable delay times and high retrieval efficiency, another figure of merit often considered critical for short- to intermediate-scale quantum memories is the time-bandwidth product (TBP). It is defined as
\begin{align}
    \label{Eq:TBP}
    TBP = \Delta \nu \Delta T,
\end{align}
where $\Delta \nu$ is the bandwidth and $\Delta T$ is the storage time. 

A large TBP enables the system to store short-duration, broadband optical pulses, making it ideal for high-speed communication. It allows for the buffering of multiple temporal modes, which is essential for synchronization, temporal multiplexing, and feedforward operations in quantum networks. 

The time-bandwidth performance of the system is primarily governed by the reflectance of the broadband optical coatings. This enables compatibility with a wide spectral range, distinguishing it from other memory approaches where the bandwidth is often tied to material properties. The broadband dielectric coatings on the delay line mirrors are designed to provide a reflectance exceeding $99.99\%$ over a $\pm30$~nm spectral range centered at $565$~nm. This was verified by measuring the retrieval efficiency of the delay line at nine different wavelengths between $532$~nm and $595$~nm, with the delay set to $370$~ns (204 reflections). The manufacturer's model for reflectance and measured reflectance at select wavelengths are presented in Fig.~\ref{fig:Measured_Reflectance}. The data demonstrates an average reflectance greater than $99.98\%$, with slight deviations attributed to scattering from dust particles and the atmosphere.

The $60$~nm bandwidth centered at $565$~nm results in a frequency bandwidth of $56.4$~THz leading to TBP = $3.87 \times 10^7$, one of the highest reported for quantum memories to date.

\subsection{Preservation of Polarization Entanglement}
Preserving the polarization state of single photons is essential for polarization-encoded quantum communication. To evaluate this, quantum process tomography and two-photon quantum state tomography were performed.

The action of the delay line on an input state $\rho_{in}$ of a single qubit can be represented by
\begin{equation}
\label{chiMatrix}
\rho_{out} = \sum_{m,n = 0}^3 \chi_{mn}E_m\rho_{in}E_n^{\dagger},
\end{equation}
where $\chi$ is the error correlation matrix \cite{Chuang1997} for the delay line. The set $\lbrace E_{m} , E_{n}\rbrace$ forms the basis of operators. An ideal channel is the identity channel, for which the $\chi$-matrix consists of only the identity element. In the case of a single qubit, a commonly used operator basis is the Pauli basis, where $E_0 = I, E_1 = X, E_2 = Y$, and $E_3 = Z$. This $4 \times 4$  Hermitian $\chi$-matrix fully characterizes the quantum channel.
\begin{figure}[h!]
\centering
\includegraphics[scale=0.06]{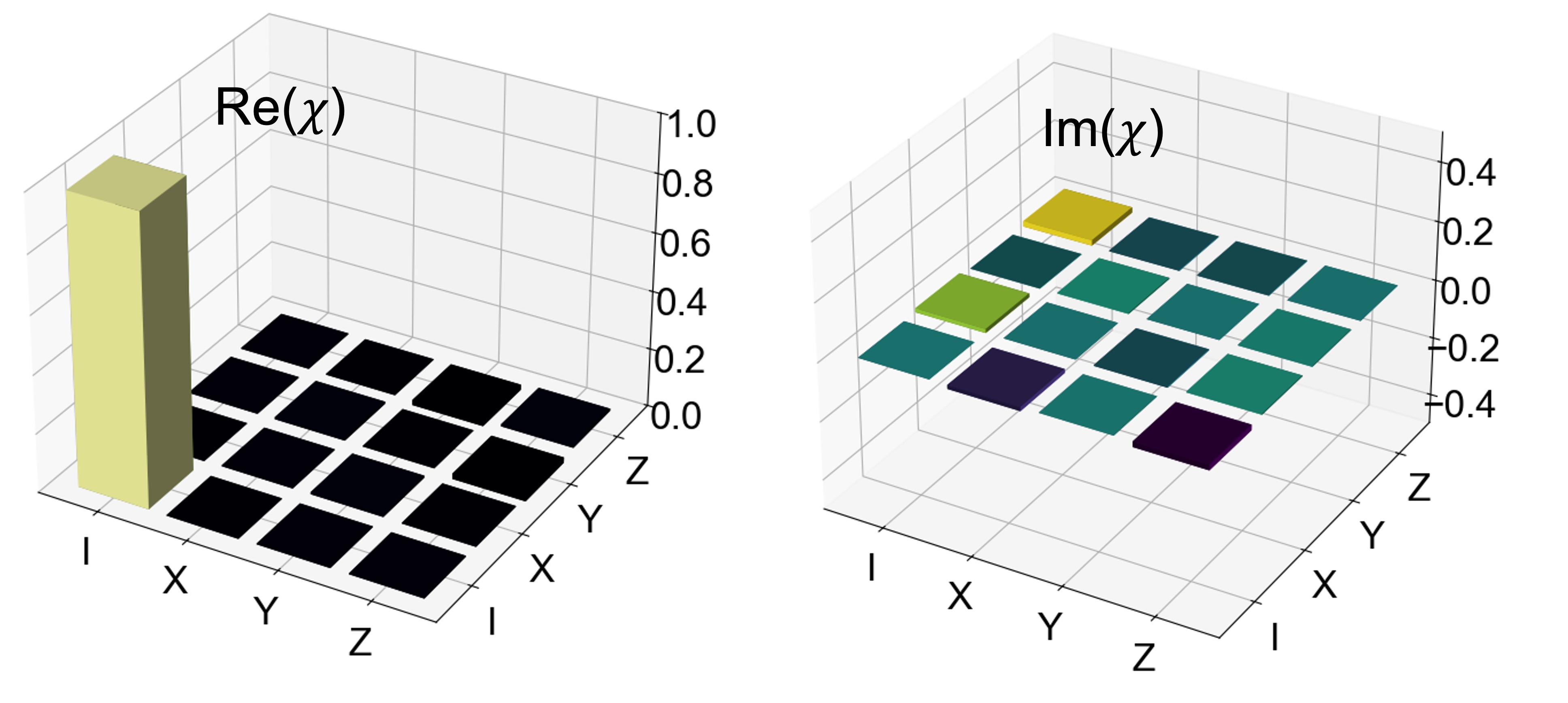}
\caption{Recreated $\chi$-matrix of the optical delay line. It has a $99.1(5)~\%$ fidelity with an identity quantum channel for polarization qubits.}
\label{fig:QPT}
\end{figure}
In the experiment, process tomography was implemented in the polarization basis. Six polarization states ($\vert H \rangle$, $\vert V \rangle$), ($\vert D \rangle$, $\vert A \rangle$), and ($\vert R \rangle$, $\vert L \rangle$) forming the eigenstates of the Pauli operators $Z$, $X$, and $Y$ respectively, were used as input states. Quantum state tomography was performed on the output states, allowing the reconstruction of the $\chi$-matrix. The results are presented in Fig. \ref{fig:QPT}. The reconstructed $\chi$-matrix exhibits a fidelity of $99.1(5)\%$ with the identity channel confirming polarization-independent performance of the delay line.

Next, polarization visibility analysis and two-photon quantum state tomography was conducted using entangled photon pairs. Coincidence counts across 16 polarization settings were used to reconstruct the density matrices before and after the delay, using standard methods \cite{James2001, White2001, Munro2001}.

\begin{figure}[h!]
\centering
\includegraphics[width=\textwidth]{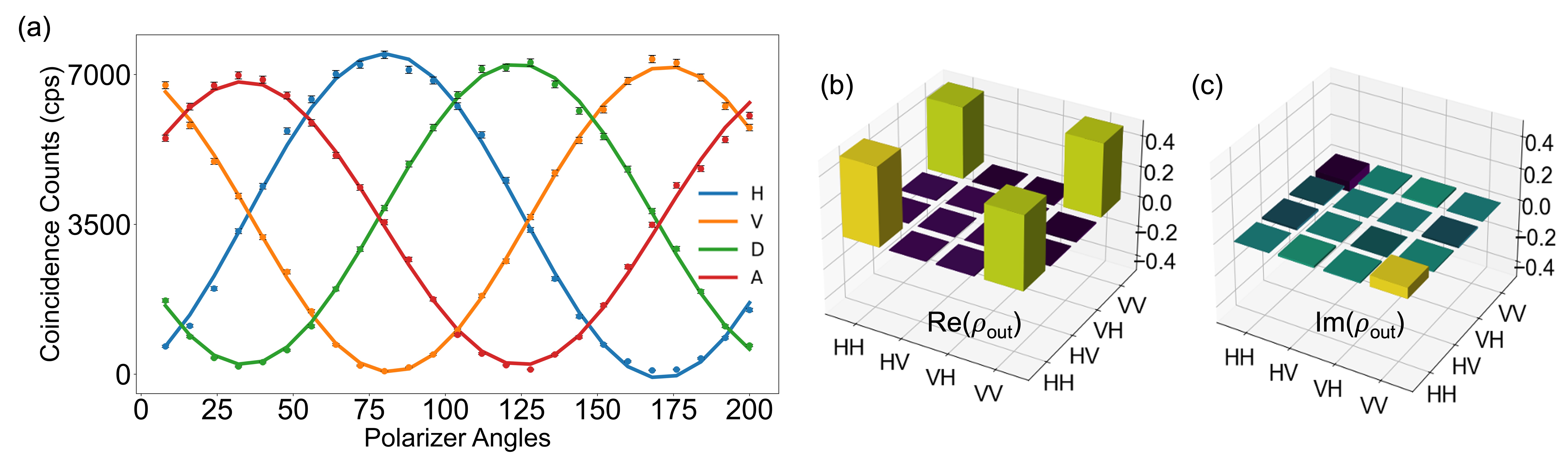}
\caption{Entanglement preservation after the signal photons have transited the delay line. (a) Polarization correlation within photon pairs in two different bases. (b) and (c) Reconstructed density matrix of the two-photon state after retrieving the signal photon from the delay line. It has a fidelity of $99.6(9)~\%$ with the input state.}
\label{fig:MemoryFidelity}
\end{figure}

The measurements are presented in Fig. \ref{fig:MemoryFidelity}, revealing high polarization visibility in both H/V and D/A basis. Without the delay line, the source exhibited an average visibility of $97.65(1)\%$. After routing the signal through the delay line and retrieval, the average visibility remained high at $97.50(2)\%$ (shown in Fig. \ref{fig:MemoryFidelity}(a)), indicating negligible degradation in polarization visibility in the delay line. These results confirm that polarization entanglement is well-preserved through the delay process. This conclusion is further supported by comparison of the input and output density matrices. The fidelity is defined as

\begin{align}
    \label{Eq:Fidelity}
    F\left(\rho_{in}, \rho_{out}\right) = \left( Tr \left[\sqrt{\sqrt{\rho_{in}} \rho_{out} \sqrt{\rho_{in}}}\right]\right)^2,
\end{align}
where $Tr$ stands for trace. $\rho_{in}$, and $\rho_{out}$ are the input and output density matrices respectively. The results of output quantum state tomography are presented in Fig. \ref{fig:MemoryFidelity}(b) and (c) which are the real and imaginary parts of $\rho_{out}$. The measured fidelity between input and output states is  $F\left(\rho_{in}, \rho_{out}\right) = 0.996(9)$.

\subsection{Comparison with delay line-based quantum memories}
\begin{figure}[h!]
\centering
\includegraphics[width=0.75\textwidth]{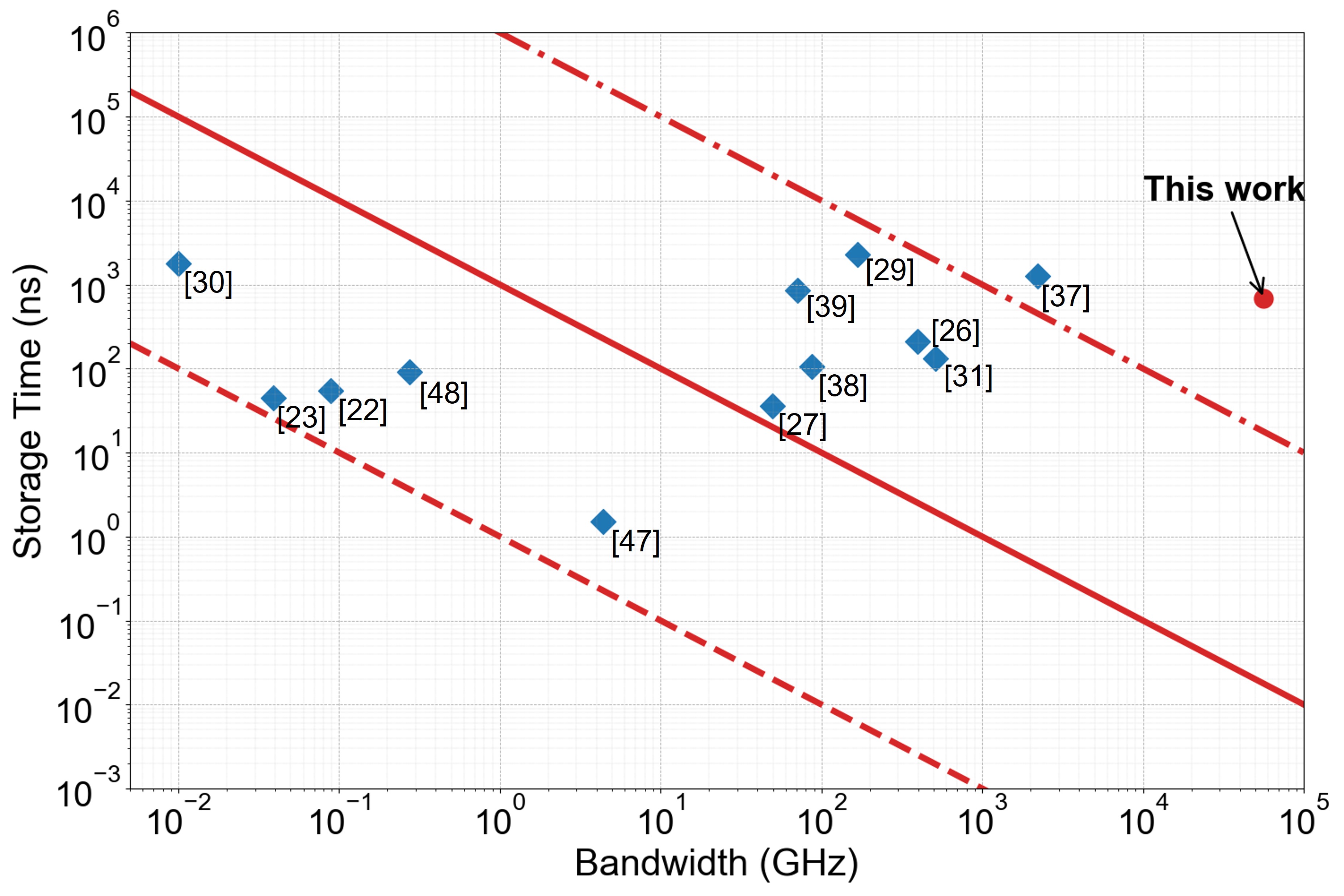}
\caption{Comparison of existing delay line-based quantum memories in terms of storage lifetime as a function of the operating bandwidth. This work reports a storage time of $687$~ns at an operating bandwidth of $56.4$~THz. Tagged references appear in the bibliography.}\label{fig:TBP}
\end{figure}

\begin{figure}[h!]
\centering
\includegraphics[width=0.75\textwidth]{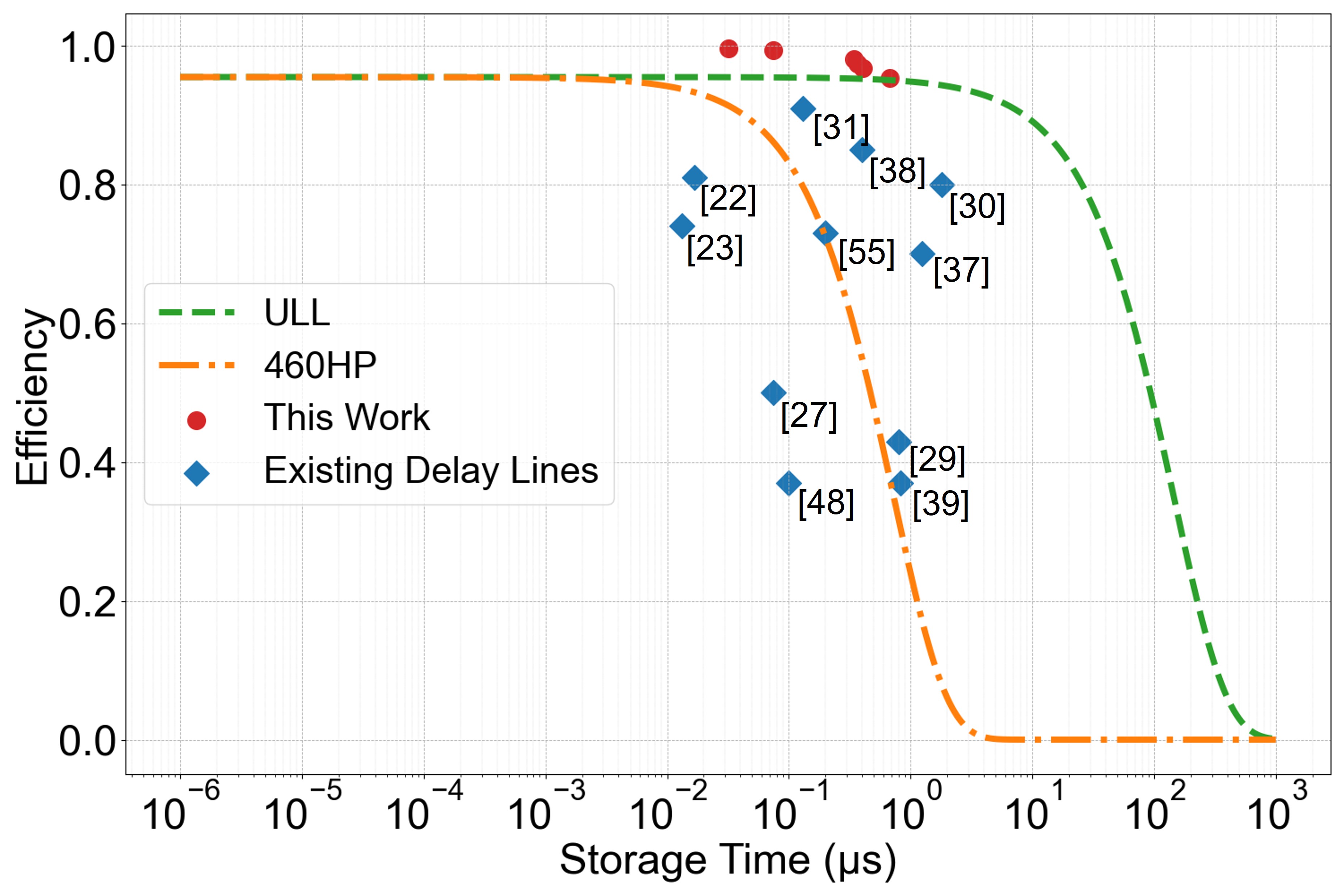}
\caption{Comparison of storage efficiency versus storage time of our work with existing delay line-based implementations. The blue curve is obtained for an ideal ultra low loss (ULL) optical fiber operating in the telecom band ($-0.15$ dB/km loss) with an additional $-0.2$ dB loss due to non-ideal coupling into the fiber and Fresnel reflection losses from the end facets. The yellow curve represents the corresponding performance obtainable with a 460HP optical fiber which is compatible with our signal wavelength. Tagged references appear in the bibliography.}\label{fig:QM_comparison}
\end{figure}

To contextualize these results, the performance of this nested multipass cell is compared with existing quantum memories based on delay lines, across two metrics: storage time vs. bandwidth (Fig. \ref{fig:TBP}) and storage efficiency vs. storage time (Fig. \ref{fig:QM_comparison}).

Most matter-based memories exhibit long storage times but are limited to narrow bandwidths, resulting in TBPs of order 1–1000 (see Fig. 10 in \cite{elizabeth2023}). In contrast, delay line-based quantum memories can typically function at higher bandwidths, with their TBP reaching $10^6$. In comparison, the nested multipass cell presented in this work achieves a TBP of $3.87\times10^7$, with a delay of $687$~ns and operating in the THz-bandwidth regime—a region where few quantum memories operate efficiently. This is one of the highest values reported even among delay line-based memories (Fig. \ref{fig:TBP}).

The efficiency vs storage time of this work is compared to other delay lines (see Fig. \ref{fig:QM_comparison}). A very common delay line is the optical fiber. For a wavelength of $565$~nm which is in the center of the operating spectrum of the nested multipass cell, a suitable optical fiber is the 460HP. At $565$~nm, the fiber introduces a loss of $\sim60\%$ for a delay of $687$~ns.  In comparison, the multipass cell has a loss of only $4.61\%$. 

It is straightforward to obtain another set of cell coated to operate at telecom wavelengths. For operation in the telecom wavelengths, ultra low loss (ULL) fibers are available with very low attenuation, typically in the range of $-0.15$~dB to $-0.20$~dB per kilometer. Making a conservative estimate of about $5\%$ coupling loss per fiber, Fig. \ref{fig:QM_comparison} shows that our system achieves higher efficiencies. This would be a first for a free space delay line to outperform ULL optical fibers. 

\section{Methods}\label{sec: method}

\begin{figure}[h]
    \centering
    \includegraphics[width=\textwidth]{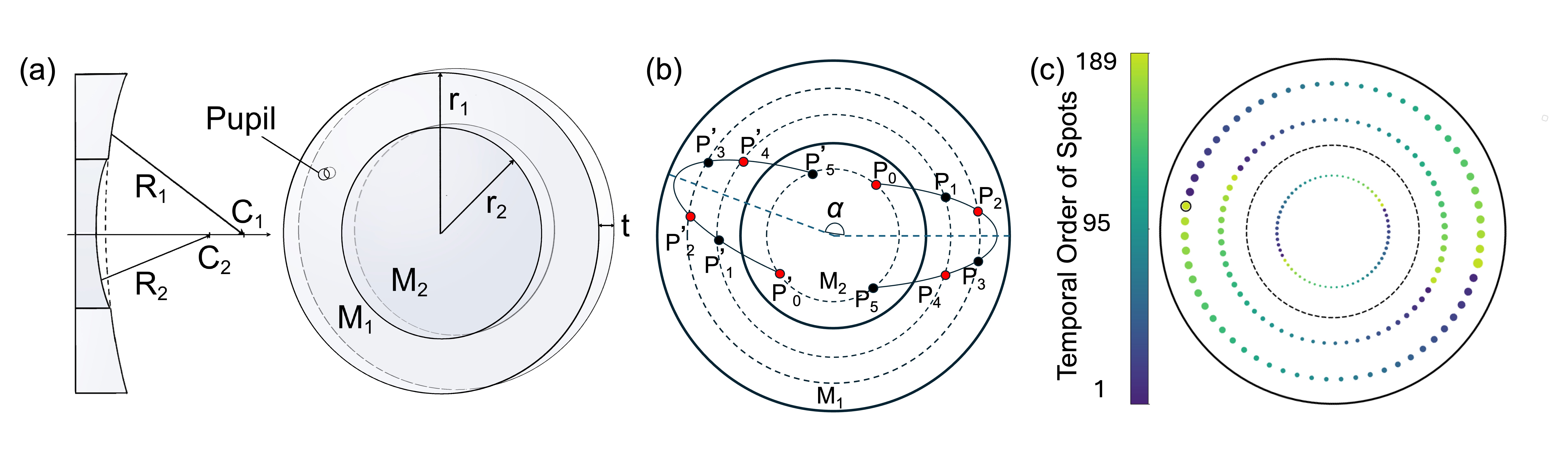}
    \caption{Geometry and reflection pattern of the nested mirror-based optical delay line.
(a) Schematic of the nested mirror configuration. A smaller concave mirror ($M_{2}$) with radius of curvature $R_{2}$ and radius $r_{2}$ is embedded coaxially into a larger mirror ($M_{1}$) of radius of curvature $R_{1}$ and radius $r_{1}$. $C_{1}$ and $C_{2}$ are the centers of curvature for $M_{1}$ and $M_{2}$ respectively. A pupil is drilled into $M_{1}$ for beam entry and exit. (b) Projection of the reflection points onto the surfaces of the two opposing nested mirrors. Red markers indicate reflections on the first nested mirror, through which the beam enters the system. Black markers correspond to reflections on the second nested mirror. The initial semi-elliptical trajectory is formed by the sequence $P_{0}$ to $P_{5}$. At $P_{5}$, the beam reflects from the second mirror and begins a new semi-ellipse, starting at $P'_0$, rotated by an angle $\alpha$ due to the curvature mismatch. (c) Simulated reflection spot pattern on the exit mirror. The color gradient from blue to green indicates the temporal order in which the spots are generated. The size of the dots are derived from Gaussian beam simulations and represents the spot size at each reflection point.}
\label{fig:MirrorDesign}
\end{figure}

The nested multipass cell configuration implemented in this work is based on the concentric Herriott cell design \cite{xue2023generalized}, which generalizes the conventional re-entrant geometry of the standard Herriott cell \cite{herriott1964off}. In this design, each nested mirror consists of a smaller concave mirror embedded coaxially within a larger spherical mirror. The two mirrors are fabricated with distinct radii of curvature and share a common optical axis. A $3$ mm circular hole (pupil) is drilled into the outer mirror to allow beam to enter and exit. A schematic of the geometry of the nested mirror is shown in Fig.~\ref{fig:MirrorDesign}(a). The larger concave mirror, $M_{1}$, has a radius of curvature $R_{1}$ and an aperture radius $r_{1}$, while the smaller mirror, $M_{2}$, has $R_{2}$ and $r_{2}$, respectively. 

A pair of nested mirrors is placed coaxially, with a separation distance $d$. After a beam enters the system through an entry pupil, its reflection spots start tracing out an elliptical path until it reflects off the inner mirror (e.g. point $P_{5}$ in Fig. \ref{fig:MirrorDesign}(b)). At this point, it encounters a different radius of curvature $R_{2}$ due to which it is reflected to point $P'_{0}$ and starts to trace out a new elliptical path. This process repeats with each subsequent interaction with the inner mirror $M_{2}$, resulting in a sequence of semi-elliptical orbits, each offset by an angle $\alpha$. 

As the beam propagates through the system, these rotated trajectories collectively form a concentric ring pattern of reflection spots on the mirror surface. In our configuration, the pattern comprises 2 rings on $M_1$ and 1 ring on $M_2$. The beam is coupled out at the spot that overlaps with the exit pupil on the outermost ring. A simulated formation of concentric spots pattern is shown in Fig. \ref{fig:MirrorDesign}(c).

The delay time of the nested multipass cell-based optical delay line depends on the number of reflections the beam undergoes before exiting the system. In this work, every $6^{th}$ reflection spot always falls on the outermost ring on which the exit pupil is positioned. The beam then couples out of the system during the $(6i+1)^{th}$ ($i = 0,1...,63$) pass, resulting in an optical path length of $\sim(6i+1)d$.

To control the number of reflections spots, the exit mirror is rotated about its optical axis to reposition the exit pupil along the outermost ring. This rotation selects which spot falls onto the pupil and determines when the beam exits the system. This approach provides controllable delay time as shown previously in Sec. \ref{sec:Result}.

The detailed parameters of the nested multipass cell used in this work are summarized in Table~\ref{tab:mirrorprameters}. These parameters were chosen to optimize the number of reflections, beam confinement, and alignment tolerance within a compact optical setup. The mirrors are mounted on custom rotation mounts with three adjusters for fine alignment control, ensuring stable and reproducible spot patterns. To achieve high storage efficiency, custom coatings were applied to all mirror surfaces. These broadband high-reflectance coatings were designed to exceed $99.99\%$ reflectance over a $60$~nm bandwidth centered at $565$~nm, improving the overall performance as shown in Sec. \ref{sec:Result}.

\begin{table}[h]
\centering
\begin{tabular}{|l|c|c|c|c|c|c|c|}
\hline
\textbf{Parameters} & $r_{1}$ (mm) & $r_{2}$ (mm) & $R_{1}$ (mm) & $R_{2}$ (mm)  &$t$ (mm) & $d$ (mm) & $V$ (L)\\
\hline
\textbf{Values} & 25.4 & 50.8 & 3355 & 4000 & 12.7 & 541 & 4.39\\
\hline
\end{tabular}
\caption{Nested cell parameters. $r_{1}$ and $r_{2}$: Radius of the outer and inner mirror; $R_{1}$ and $R_{2}$: Radius of curvature of outer and inner mirror;  $t$: Thickness of mirror; $d$: Separation distance between two nested mirrors; $V$: Volume of the system.}
\label{tab:mirrorprameters}
\end{table}

\section{Discussion}\label{sec: conclusion}
This work demonstrats a free space optical delay line-based on a nested multipass mirror configuration, capable of preserving quantum polarization states with high fidelity and efficiency. The device supports discretely tunable delays up to $687$~ns in steps of $\sim12.6$ ~ns, maintains a retrieval efficiency of $95.390(5)\%$, and achieves a fidelity of $99.6(9)\%$ for entangled photon states. The resulting time-bandwidth product of $3.87 \times 10^{7}$ exceeds that of most reported platforms.

The distinctive advantages of the nested multipass cell lie in its operational simplicity and versatility. The system operates entirely at room temperature, does not rely on active control, and supports broadband operation without the constraints of narrow spectral resonances. These features reduce technical complexity and can potentially store qubits encoded in different degrees of freedom. The ability to preserve polarization entanglement throughout the entire storage and retrieval process highlights its potential as a high-fidelity quantum memory. In addition, the exceptionally large time-bandwidth product ($\sim 10^{7}$) surpasses that of most reported implementations, offering substantial capacity for future photonic quantum storage and synchronization protocols.

Looking ahead, integrating fast optical switches \cite{pengfei2024} into the delay line could enable dynamic control over photon routing, allowing multiple sequential transits and on-demand retrieval. This would transform the current system into a fully programmable all-optical quantum memory, with storage times extendable to the microsecond regime while preserving high efficiency. Furthermore, the broadband nature of the system ensures robustness against spectral variations, supporting compatibility with a wide range of photonic sources.

This work opens the door to scalable and programmable all-optical quantum memories that operate at room temperature and require no material excitation or complex controls. Requiring less stringent operational conditions and exhibiting compatibility with standard photonic components, the system establishes the basis for flexible, on-demand quantum storage enabled by future integration with optical switching technologies.

\section*{Acknowledgements}
This project is supported by the National Research Foundation, Singapore through the National Quantum Office, hosted in A*STAR, under its Centre for Quantum Technologies Funding Initiative (S24Q2d0009), and the Ministry of Education, Singapore under its Tier 3 grant SENIOR (MOE-MOET32024-0002).  The authors would like to acknowledge Paul Kwiat and Nathan Arnold for insightful discussions, and Grace Pang for supporting the stylization of some of the figures presented here.

\begin{appendices}

\end{appendices}


\end{document}